\documentclass[draft]{agujournal2019}
\usepackage{url} 
\usepackage{hyperref}
\usepackage{soul} 

\usepackage{anyfontsize}
\usepackage{xcolor}
\usepackage[utf8]{inputenc}
\usepackage{subcaption,setspace,graphicx} 
\usepackage{gensymb}
\usepackage{bm,amsmath}  
\usepackage{natbib}
\usepackage{newtxtext,newtxmath}
\usepackage{booktabs,array}
\usepackage{multirow,makecell}  

\draftfalse

\journalname{JGR: Planets}

\begin{document}

%
%

\title{Early solar wind and dynamo magnetic field topology predictions for (16) Psyche and other asteroids}


%
%




\authors{Atma Anand\affil{1}, Jonathan Carroll-Nellenback\affil{1,3}, Eric G.~Blackman\affil{1,3}, John A.~Tarduno\affil{2,1,3}}


\affiliation{1}{Department of Physics and Astronomy, University of Rochester, Rochester NY 14627}
\affiliation{2}{Department of Earth and Environmental Sciences, University of Rochester, Rochester NY 14627}
\affiliation{3}{Laboratory for Laser Energetics, University of Rochester, Rochester, NY 14623, USA}




\correspondingauthor{Atma Anand}{atma.anand@rochester.edu}




\begin{keypoints}

\item If a dipolar field aligned with asteroid’s spin axis is observed, magnetohydrodynamic modeling suggests magnetization by early solar wind

\item A dipolar field tilted from the asteroid’s spin axis would imply it is a fragment of an impact disrupted body magnetized by a core dynamo

\item A highly non-dipolar field would suggest that the asteroid is a largely intact body magnetized by an internal dynamo or the solar nebula

\end{keypoints}

%
%

%
%


\begin{abstract}
Asteroid (16) Psyche is a metal-rich body that might record an ancient coherent magnetization if some relict crust or mantle is preserved. Herein, we use magnetohydrodynamic simulations to predict (16) Psyche’s field topology for several distinct pathways: (i) an early solar wind-induced magnetization imparted after a larger body was impacted, forming the present-day asteroid, (ii) a core dynamo magnetization imparted in an asteroid that is either presently largely intact or was a rubble pile, and (iii) magnetization in the turbulent solar nebula disk. For pathway (i) we find the field to be predominantly dipolar and spin axis-aligned. For pathway (ii) we find the field to be either dipolar and spin axis-misaligned, or highly multipolar. We also find that (iii) a field produced earlier before the solar nebula cleared, would be highly multipolar. In cases (i) and (ii) we also place constraints on the field strength. Simple detection of a magnetic field without constraining its topology and temporal variability would be insufficient to confirm a remanent source, due to the influence of the present-day solar wind, electromagnetic induction, and (16) Psyche's high obliquity. For sufficiently strong fields however, the field topology and orientation may reveal key observable consequences of the nature and history of (16) Psyche. Our framework is also broadly applicable to the study of magnetic fields from other asteroids.

\end{abstract}

\section*{Plain Language Summary}
Magnetic fields sensed by satellites can provide key information on the nature of planetary bodies. Here we use computational modeling to predict the magnetic fields that could be sensed by NASA’s mission to (16) Psyche, a metal-rich body in the asteroid belt. For a sufficiently strong magnetic field, the specific strength, shape, and orientation of the field relative to the spin of (16) Psyche could discriminate whether Psyche once had a liquid iron core generating a magnetic field, whether the  body is  intact or was fragmented, or whether magnetic fields associated with intense solar winds imparted a magnetic signal to the asteroid after it was impacted early in the history of the Solar System. {However, mere detection of a weak  magnetic field value would be inconclusive as to its source because the solar wind's interaction with (16) Psyche can create confounding magnetic signatures. This method can be generalized to study the magnetic fields of other asteroids. 

%
%

\section{Introduction}
M-class asteroids were classically interpreted as potential fossil cores of larger planetary bodies because their visible/near-infrared spectra suggested compositional similarities with iron meteorites \citep{Bell1989}. With an average diameter of 233 km, (16) Psyche is the largest M-class asteroid and a natural target for exploration \citep{Elkins-Tanton+2020}.  However, more recent imaging and analyses indicate that (16) Psyche’s density of $4160 \pm 640 \, \mathrm{kg \, m}^{-3}$ \citep{Drummond2018}, or $3880 \pm 250 \, \mathrm{kg \, m}^{-3}$ \citep{Siltala2021} is lower than that of iron meteorites, strongly suggesting that while still metal rich, (16) Psyche has a significant porosity and/or is a metal-silicate mixture.  This finding raises fascinating possibilities for the nature and evolution of (16) Psyche. For example, it is possible that (16) Psyche formed with a very thin mantle, discussed as one possibility for the nature of the Main Group pallasite meteorite parent body \citep{tarduno2012,nichols2021}. 
Alternatively, an impact may have stripped (16) Psyche of most of its mantle, leaving a body with some mantle over an intact or disrupted core \citep[e.g.,][]{Asphaug2006}. Such differentiated iron-rich bodies can result from hit-and-run collisions during planetary formation \citep{Asphaug2014}. Metal-rich bodies may also form directly from material with high metal-to-silicate ratios set by the reducing initial conditions of their local accretion environment \citep{Suer2025}.
These scenarios further raise the question of how magnetometer data obtained during NASA’s current mission \citep{Elkins-Tanton+2020,Weiss2023} might help distinguish between alternatives for the nature and history experienced by (16) Psyche, a topic we take up anew here.

Following pioneering work on meteorite magnetism, it has been recognized that Solar System bodies can be magnetized by external processes or internal processes \citep{Banerjee1972,nagata1979}. External processes include direct magnetization by nebular \citep[e.g.][]{Fu2020,Bryson2023,Mansbach2024} or solar wind fields \citep{obrein20}. In addition, impacts are an external process that can magnetize bodies by impact plume-driven compression of the ambient solar wind field \citep{Hood1987} or charge separation in the impact plasma \citep{Crawford2020}. 
Internal processes usually invoke a core dynamo \citep[e.g.,][]{Stanley2014,Nimmo2009}, though for asteroids and planetesimals, thermal and compositional convection in metallic cores can sustain dynamos on timescales of millions to tens of millions of years \citep{Zhang2023}. For large terrestrial planetary bodies, mantle magma ocean dynamos are also possible \citep[e.g.,][]{Stixrude2020} , but this  is not applicable to asteroid-sized bodies and is not considered further here.
A third, distinct mechanism involves local, grain-scale self-magnetization, where the characteristic of some FeNi magnetic minerals, particularly those of iron meteorites, allows them to acquire a remanent magnetization when cooled in a zero magnetic field \citep{Brecher1977} and/or by interactions between adjacent magnetic minerals \citep{dunlop1997rock,obrein20}.

Faced with this plethora of magnetization pathways, here we primarily consider three distinct pathways:
(i) Early solar Wind-Induced Magnetization (WIM): The present-day (16) Psyche formed after impact disruption of a larger differentiated body. The impact stripped most but not all of the parent body's mantle, heating the remaining material above the blocking temperature of FeNi magnetic minerals ($\sim 360$ to $760\degree$C, \citealt{dunlop1997rock}). Upon cooling in the ram pressure-amplified solar wind field of the early Solar System, these minerals acquired a thermoremanent magnetization \citep{obrein20,Anand2021}.
(ii) Core dynamo magnetization: We investigate two distinct dynamo scenarios. First, (16) Psyche could be a largely intact differentiated body with an appreciable silicate mantle that recorded a magnetic field from its own past core dynamo. Second, (16) Psyche could be a small fragmented rubble pile from the mantle of a larger disrupted differentiated body, preserving magnetization imparted by the parent body's core dynamo before disruption.
(iii) Solar nebula magnetization: As the formation epoch of (16) Psyche is uncertain, we also obtain preliminary results for magnetization in the presence of the solar nebula disk, applying turbulent disk models \citep{Balbus1998,Blackman2001} with multiple field reversals during the cooling timescale. However, for reasons discussed in our highlight of future directions (Section 5.4), these results are not the primary focus of this work.

Assumptions about the magnetized material are essential because iron meteorites are notoriously poor magnetic recorders \citep{nagata1978}. Any ancient magnetizations held by huge multidomain and low coercivity kamacite grains common in iron meteorites could have relaxed in the billions of years since the early Solar System. In contrast, much smaller single domain FeNi grains found in a silicate mantle can retain magnetizations on multi-billion-year timescales \citep{tarduno2012}. In this sense, the recognition of a density for (16) Psyche lower than that of iron meteorites enhances the possibility that a diagnostic magnetic signature will be recorded. If (16) Psyche were solely a very large iron body, any remanent magnetization would be questionable. We will revisit these magnetization assumptions, as well as the age of magnetization relative to dispersal of the solar nebular, in our discussion of results.

We carry out magnetohydrodynamic calculations using the adaptive mesh refinement (AMR) code AstroBEAR \citep{cunningham09,carroll13}, to better understand the magnetization scenarios (i-ii) described above. \cite{obrein20} and \cite{Anand2021} previously used this code to investigate the magnetizing potential of the early solar wind, shortly after the clearing of dust and gas of the solar nebula. When encountering a spherical carbonaceous chondrite asteroid, the field was found to be amplified to values $ \sim 1$ µT for bodies at orbital radii equal to or less than those of the present-day asteroid belt. While the viability of solar wind magnetization was questioned based on a limited parameter exploration \citep{oran18}, a more comprehensive analysis  across a broader parameter space demonstrates viable magnetization pathways \citep{obrein20,Anand2021} under early solar system conditions \citep[e.g. for the solar wind,][]{Vidotto2021}. The case considered by \cite{obrein20} and \cite{Anand2021} addressed the potential to magnetize, on time scales less than a solar reversal cycle period, a near surface sample represented by a meteorite, and magnetic minerals that formed by metasomatic reactions common on early carbonaceous chondrite parent bodies. Hence, this magnetization is a low temperature chemical or crystallization remanent magnetization.

Here, we generalize this prior application of  wind induced magnetization (case (i) above),  to a biaxial ellipsoid, which is a better approximation for (16) Psyche \citep{Shepard2021}. Like the previous work, we use the amplified magnetic field from the simulations as the externally applied field to the body, but we now also model the consequent magnetization of the surface of (16) Psyche over various acquisition time scales to represent the thermoremanent magnetization of the entire body whose external field might be sensed by a satellite. For comparison, we use a similar procedure to predict the remanent field that would arise from (a) an internal or (b) parent body core dynamos (case (ii) above). 

\section{Methods: Parameters for (16) Psyche Simulations}
We describe our simulation setup, post-processing procedure, choice of physical parameters, and the physics we consider in each step.

\subsection{Asteroid Model and Simulation Setup}
The best fit ellipsoid for (16) Psyche has axes of 139 km $\times$ 119 km $\times$ 85.5 km \citep{Shepard2021}. Since the larger two axes are close in size, for simplicity we volumetrically approximate the biaxial dimension using the known volume of (16) Psyche, $(5.75 \pm 0.19) \times 10^6 \text{ km}^3$ \citep{Shepard2021}, and 85 km for the smaller (spin) axis. The result is rounded \textit{down} to a biaxial ellipsoid of 125 km $\times$ 125 km $\times$ 85 km for our wind induced magnetization simulations. (16) Psyche currently has an obliquity (spin-axis tilt relative to its orbital plane) of $98\degree$. This inclination could reflect the last major collision it experienced, stripping the mantle and resulting in its magnetization as discussed above. We consider this case in P1-7, as well as generic cases G1-7 for when (16) Psyche spin-orbit inclination was not at its present value (Fig. \ref{fig:setupB} and Table \ref{tab:all}). 
These cases span the positive octant of the sphere with symmetry transformations allowing us to infer the remaining octants. To capture these generic orientations (Cases G1-7) in our resistive MHD AstroBEAR simulations, the wind (x-axis) and magnetic field (z-axis) directions are kept fixed and orthogonal to each other.
We define the spin-axis orientation using standard spherical angles: $\theta$, the polar angle from the magnetic field (z-axis), and $\varphi$, the azimuthal angle measured counter-clockwise from the wind direction (x-axis) in the x-y plane.
Each of the 7 cases examined (Fig. \ref{fig:setupB}a) corresponds to a rotated ellipsoid spin axis, which is fixed in the body frame of reference along the smallest axis (85 km, indicated by the long arrow in Fig. \ref{fig:setupB}a).

These 7 simulations in the positive octant represent three distinct SO(3) symmetry classes that, through reflection, generate a full set of 26 unique orientations spanning the unit sphere. Three of the cases (G1, G5, G7) lie on the coordinate axes, and reflections generate 2 unique orientations for each, for a total of 6 orientations. Three other cases (G2, G4, G6) lie on the coordinate planes, generating 4 unique orientations each, for a total of 12 orientations. The final G3 case uses $\theta = \arccos(1/\sqrt{3}) \approx 54.74\degree$. This specific orientation places the spin axis equidistant from all three coordinate axes (x, y, z), representing a case of maximal asymmetry relative to our orthogonal wind-field setup and generates 8 unique orientations under SO(3) symmetry transformations, one for each octant. The sum of these symmetrically-generated orientations is $6 + 12 + 8 = 26$. These were used to model the multiple turbulent eddy crossings for the solar nebula magnetic acquisition case (Supplementary Fig. S2) without running additional simulations.

\subsection{Magnetization Physics and Parameters}
The net dipole moment of the asteroid is:
\begin{equation}
\rm{M_p} = \frac{ \chi_M \, \rm{B} \, 4 \pi R_p^2 d}{ \mu_0 },
\label{dipole}
\end{equation}
where, $M_P$ is the asteroid's magnetic dipole moment, $\chi_M$ is the dimensionless (density accounted) magnetic susceptibility, $R_p$ is the volumetrically averaged radius of (16) Psyche, 111 km, $\mu_0$ is the magnetic permeability of free space, $d \ (<<R_p)$ is the thickness of the crust that gets magnetized, and  \rm{B} is the magnetic flux density (magnetic field) at the surface from the compressed solar wind.
For a given \rm{B} and $\chi_{M}$, the required magnetization depth $d$ can be inferred by combining Equation \ref{dipole} with the solar wind pressure-balance condition. This calculation determines the minimum $d$ necessary to produce a magnetopause standoff distance at the spacecraft's orbital altitude, which serves as a detection threshold.

We assume that an outer crust or mantle layer of thickness \textit{d} with a uniform magnetic susceptibility is magnetized by pathways (i), internal dynamo in (ii), and (iii) described earlier. We select as an upper limit a mass-normalized magnetic susceptibility, $\chi = 6$ in log units of $10^{-9} \mathrm{m}^3 \mathrm{kg}^{-1}$, bounding measurements of mesosiderite and pallasite meteorites \citep{Macke2010, MACKE2011,Uehara2023} , thought to represent metal-silicate impact mixtures \citep{Scott2001,tarduno2012,Walte2020,Windmill2022}, as well as acapulcoites and lodranites, iron rich primitive achondrites \citep{McCoy1997,Floss2000}. Using a density of $4000 \, \mathrm{kg\ m}^{-3}$ for (16) Psyche \citep[e.g.][]{Elkins-Tanton+2020} and $ \chi = 10^{-3} \mathrm{m}^3 \mathrm{kg}^{-1}$, our upper magnetic susceptibility bound yields a density-corrected magnetic susceptibility value of $\chi_M = 4$. We discuss the implications of this susceptibility bound and layer thickness in section 4.

\subsection{WIM Amplification \& MHD Applicability}
The maximum field magnitude amplification $A$ from solar wind induced magnetization is the minimum of $A \approx M\sqrt{\gamma\beta}$ and $A \approx R_{M}$ \citep{Anand2021}, where $\beta$ is the ratio of wind thermal to magnetic pressure, $\gamma$ is the adiabatic index, $M$ is the wind Mach number, and the magnetic Reynolds number $R_{M} = 2 v_w R_p / \eta$ is a ratio of the magnetic diffusion time to the wind crossing (advection) time,
where $v_w$ is the wind speed, $R_p$ is the body radius, and $\eta$ is the magnetic diffusivity. The final magnetizing field $B$ is estimated from the ram pressure balance condition, where the amplified magnetic pressure $B_{\text{amp}}^2/8\pi$ stands off the solar wind ram pressure $P_{\text{ram}} = \rho v_w^2$. This gives an amplified field of $B_{\text{amp}} \approx v_w \sqrt{8\pi\rho}$. The final field $B$ recorded by the material includes the material's own response: $B = B_{\text{amp}}(1+\chi_{M})$, leading to the expression:
\begin{equation}\label{eq:amp}
    B \approx (1+\chi_{M}) v_{w} \sqrt{8\pi\rho} = 2.1\,\mu\mathrm{T}\,(1+\chi_{M})
    \left(\frac{v_{w}}{1000\,\mathrm{km\,s}^{-1}}\right)
    \left(\frac{n}{1000\,\mathrm{cm}^{-3}}\right)^{1/2}
\end{equation}
We note that $R_{M\alpha}$ in Eq. 21 of \citealt{Anand2021} is the numerical magnetic Reynolds number, a simulation parameter related to grid resolution and numerical diffusivity, not a physical property of the asteroid. Simulations need to ensure that $R_{M\alpha} > R_M $ to avoid numerical artifacts.
For a body of (16) Psyche's estimated electrical conductivity (see Supplementary Section S2), $R_{M} \sim 1000 $. Thus, the amplification is set by the ram pressure limit, as $R_{M} \gg M\sqrt{\gamma\beta} \sim 10$.

The Larmor (or ion gyro-)radius is the characteristic radius of a charged particle's helical motion around magnetic field lines, given by $r_{\rm L} = m v_{w}/q B$, where $m$ is the ion mass (proton mass $\approx 1.67 \times 10^{-27}$ kg), $v_{w}$ is the solar wind velocity ($\sim$500 km s$^{-1}$ for the early solar wind), $q$ is the ion charge (proton charge $= 1.6 \times 10^{-19}$ C), and $B$ is the magnetic field strength ($\approx 10^{-6}$ T). For early solar wind conditions at (16) Psyche's orbital distance, this yields $r_{\rm L} \approx 10$ km. For MHD to be applicable to a collisionless plasma, the Larmor radius replaces the mean free path and must be at least an order of magnitude smaller than the characteristic system scale (here the radius of Psyche, $\sim$100 km), a condition satisfied for the early solar wind. We note that simulating present-day solar wind interactions with Psyche's remanent field requires plasma models beyond MHD, such as  hybrid codes that can separately track ions while treating electrons as a fluid
(e.g., \citealt{Fatemi2018}), as the present-day solar wind violates the MHD approximation at these scales. For more details on the MHD applicability to early solar wind magnetization, see \cite{Anand2021}. We use Wind II from \cite{Anand2021}, which has $v_w=500$ km s$^{-1}$, $B_w=100$ nT, $T=5 \times 10^5$ K, and a density of 1000 $\#$ cm$^{-3}$, which gives $M\sqrt{\gamma\beta} = 11.4$. We use a constant resistivity for (16) Psyche such that $R_M = 1500$. Supplementary S2 describes why our model is not sensitive to the exact resistivity, as long as the asteroid is conductive enough.

\subsection{Case Definitions and Post-Processing}
We define five sets of simulation cases in Table \ref{tab:all}, with case types summarized in Table \ref{tab:cases}.
To model the remanent field acquired over a rotational period, the WIM simulation results (G1--G7 and P1--P8) are post-processed via spin-averaging (see MATLAB code in Open Research Section). This is performed by first interpolating the cells in the outer 10 km of the asteroid from Cartesian to spherical coordinates. The simulated magnetic field vectors are then transformed from Cartesian into the asteroid's body-fixed spherical coordinate frame. We then average the radial, polar, and azimuthal components ($B_r, B_\theta, B_\varphi$) 
around the azimuthal (spin) axis. The vectors are then converted back into Cartesian coordinates. The resulting averaged vector components are used to calculate the final physical moments (dipole and quadrupole) for each case using volume-weighted integration (summation), as listed in Table \ref{tab:all}. 
These physical moments are then converted to the standard Gauss-Legendre coefficients ($Q_{10}=g_1^0, Q_{20}=g_2^0$, see Eqn. \ref{eqn:Vr}) reported in our table, which have units of nT. The conversion is $Q_{10} = (\mu_0 / 4\pi) (p_z / a^3)$ and $Q_{20} = (\mu_0 / 4\pi) \{(2Q_{zz} - Q_{xx} - Q_{yy}) / 2a^4\}$, where $a$ is the reference radius (111 km) and $\mathbf{p}$ and $\mathbf{Q}$ are the physical dipole and quadrupole moments.
This averaging procedure inherently cancels non-axisymmetric ($m>0$) terms, which is why the toroidal field $m_\varphi$ is non-zero in Table} \ref{tab:all} while the $m>0$ spherical modes vanish. The `PA' (Psyche Average) case is calculated subsequently to represent the average remanent field acquired over a full orbit. This can be computed by taking the simple arithmetic mean of the final moments (e.g., $Q_{10}$, $Q_{20}$, $m_z$) from the eight orbital position cases (P1--P8), or averaging the Cartesian vectors in the very first step and following all the steps. 

For the solar nebula case we generate 100 or 1000 
eddies with the \textit{peak} magnetic field on the asteroid's surface varying as a Gaussian of mean $3$ mT and variance $1.5$ mT. This estimate is based on equating the ram pressure of the turbulent nebular wind at asteroid belt distances with the magnetic pressure at the surface of the asteroid, and should be considered an upper limit. As discussed in Section 5.4, this pressure conversion is uncertain for larger bodies due to bow shock effects. The asteroid's spin axis is simultaneously sampled as a Gaussian on the unit sphere and interpolated onto one of the 26 unit sphere cases discussed above.

\subsection{Core Dynamo Magnetization Model}
Unlike the WIM cases, which are post-processed from MHD simulations, the core dynamo cases (`CP' and `CB') are modeled as idealized thermoremanent magnetization (TRM) scenarios from purely dipolar core dynamos. The `CP' (Core, Parent) case models (16) Psyche as a fragment that was fully and uniformly magnetized at the core mantle boundary of a larger parent body of size and magnetic field equal to that of present day Mercury (Fig. \ref{fig:xyAll}e).  We use Mercury's dipole moment ($M_{CP} = 2.25 \times 10^{18} \mathrm{A \, m}^{2}$) as the source analog, with the asteroid's spin axis tilted $45\degree$ relative to this external field. The `CB' (Core, Body) case models an internal dynamo that magnetizes only the outer 10 km crustal shell, identical in geometry to the WIM cases. We assume a internal dynamo with dipole moment of $M_{CB} = 10^{16} \mathrm{A \, m}^{2}$, which is inclined at $45\degree$ and offset from the asteroid's center to (25, 0, 25) km (Fig. \ref{fig:xyAll}f). In both dynamo cases, the field is a body-fixed TRM, so no spin-averaging is applied.

\begin{table}[h]
\centering
\caption{Simulation case definitions}
\label{tab:cases}
\begin{tabular}{lll}
\hline
Key & Cases & Description \\
\hline
G (Generic) & G1--G7 & Generic spin-axis orientations: on axes (G1, G5, G7), \\
 & & on planes (G2, G4, G6), and equidistant from axes (G3) \\
P (Psyche) & P1--P8 & (16) Psyche's current $98\degree$ obliquity with $\varphi$ in $45\degree$ steps \\
PA (Psyche Average) & PA & Orbital average of P1--P8 (mean of $m_x$, $m_y$, $m_z$) \\
CB (Core, Body) & CB & \textit{Internal} core dynamo from (16) Psyche itself \\
CP (Core, Parent) & CP & \textit{External} core dynamo; (16) Psyche as fragment magnetized \\
 & & by larger parent body \\
\hline
\end{tabular}
\end{table}

\begin{figure*}
    \centering
    \includegraphics[width=\textwidth]{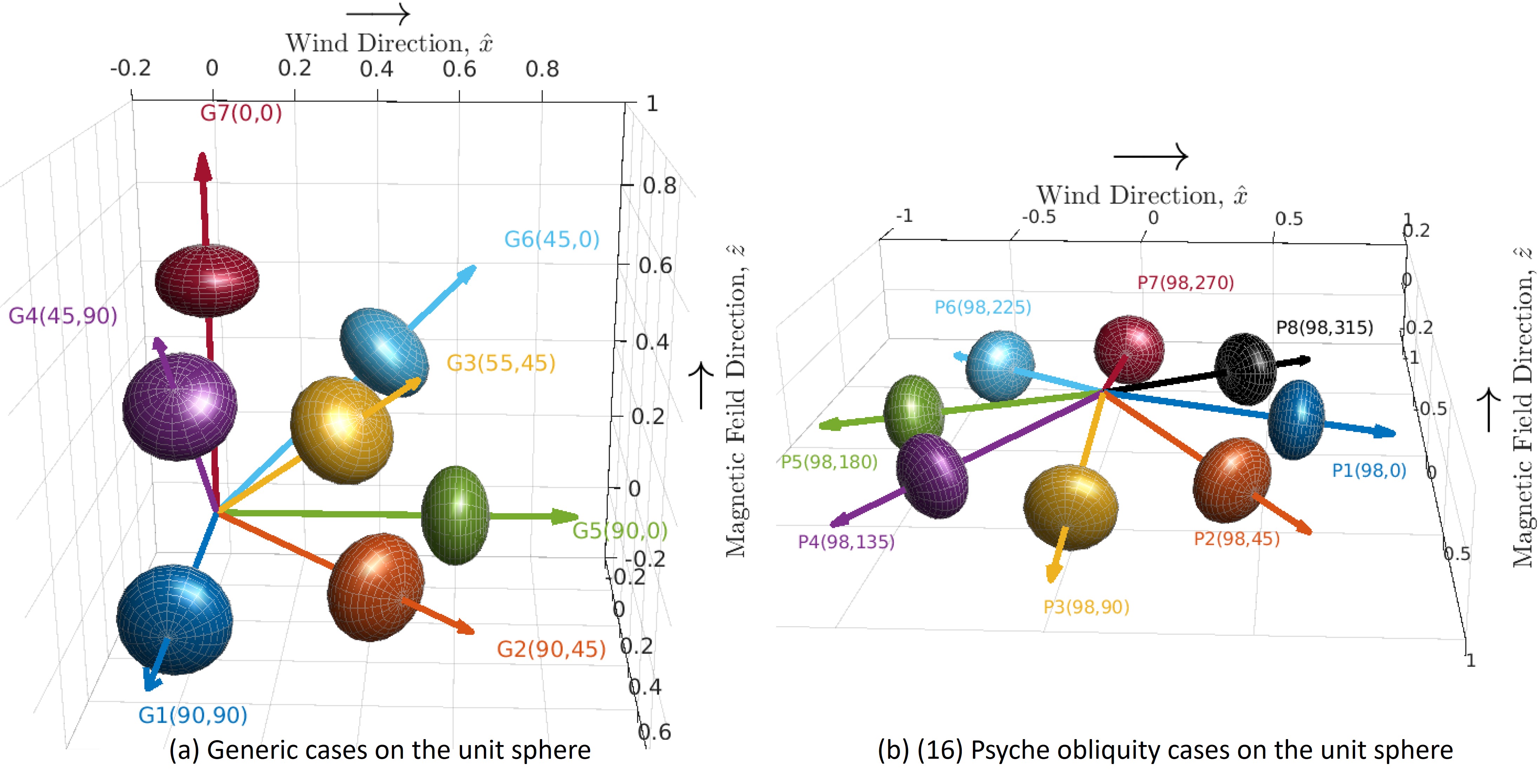}
    \caption{Our simulations are carried out in a frame centered on the ellipsoidal asteroid, but here we show all  spin axis orientation cases that we consider  in the frame of reference of the background solar wind/magnetic field. The asteroid configuration is represented by the colored solid ellipsoids with the shortest axis being the spin axis. The long arrows  represent the spin axes  for each particular simulation. The labels show the angle of the spin axis from the z-axis or magnetic field direction ($\theta$), and x-axis or wind propagation direction ($\varphi$). Panel (a) shows  cases that generically span the positive octant of spin orientations on a unit sphere, from which all other analogous octants can be determined by symmetry. Panel (b) shows cases with obliquity $(\theta)=98\degree$ and $\varphi$ at $45\degree$ increments for averaging over the orbit (revolution) period of (16) Psyche.
    }
    \label{fig:setupB}
\end{figure*}

\section{Simulation Results : Effect of Spin Orientation and Magnetizing Source on Predicted Magnetic Fields}
\begin{figure*} 
    \centering
    \includegraphics[width= \textwidth]{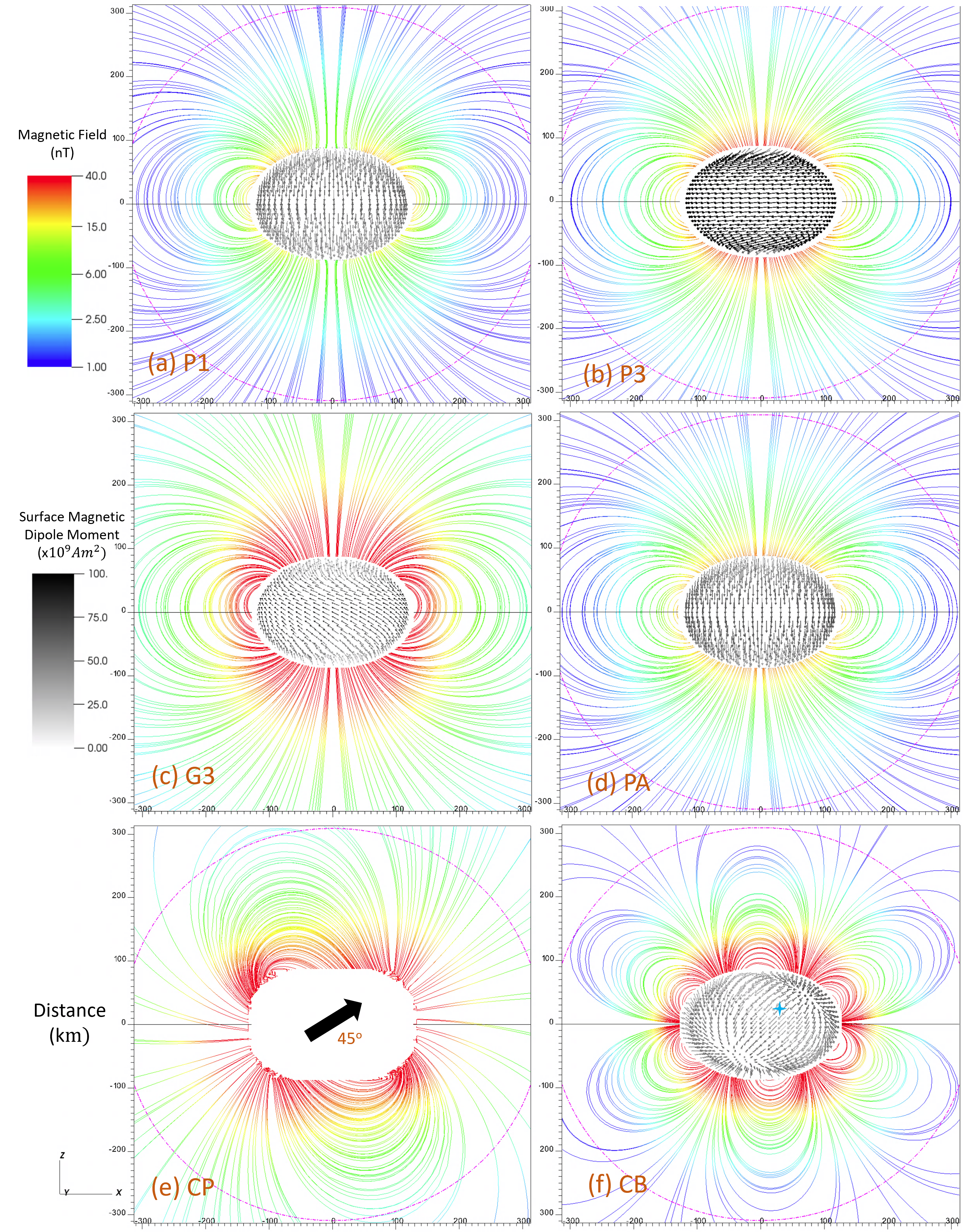}
    \caption{Synthetic remanent  magnetic field lines colored by magnitude and surface remanent  magnetization
    (grayscale) in the  $y=0$ plane. The $z=0$ line
    highlights the North - South symmetry or  asymmetry in some cases. Top and middle rows are the WIM cases after spin averaging with the spin axis oriented at (a) $ \theta = 98 \degree, \ \varphi = 0 \degree $, (b) $ \theta = 98 \degree, \ \varphi = 90 \degree $, (c) $ \theta = 54.74 \degree, \ \varphi = 45 \degree $, (d) is further averaged over present day Psyche's orbit (8 cases with $ \theta = 98 \degree$ as shown in Fig. \ref{fig:setupB}b.
     Panel (e) shows a core dynamo imparted field if Psyche were at the core mantle boundary 
     of a larger parent body with the size and magnetic field of present day Mercury with its spin axis at $45 \degree$ from the core dynamos dipole. The  surface remanent magnetization is not shown as the whole body is magnetized. Panel (f) is a case of a strong core dynamo ($10^{16} \mathrm{A \, m}^2$) inclined at $45 \degree$ in the $xz$ plane and slightly off-center of Psyche itself at (25, 0, 25) km (denoted by a blue star) and  for which the crust cooled from the outside inward. Case (f) is multipolar and non-axisymmetrical, hence our analysis is only first order accurate. On the panels most relevant for (16) Psyche, and thus the Psyche spacecraft, we have shown dot-dashed violet ellipses  representing one of several orbits the probe will take at $ 100 \ \mathrm{km} $ altitude from the asteroid's body. The labels in the bottom left of each subfigure correspond to the corresponding entries of Table \ref{tab:all}.}
    \label{fig:xyAll}
\end{figure*}

We separately analyze our AstroBEAR simulation results for the solar wind induced magnetization case and the core dynamo case (Fig. \ref{fig:xyAll}, Table \ref{tab:all}) in this section. Supplementary Figure 2 shows the case of being magnetized while cooling and encountering multiple turbulent eddies in the solar nebula, subject to a caveat about bow shocks reducing field magnitude.

\begin{figure}
\centering
\includegraphics[width= \textwidth]{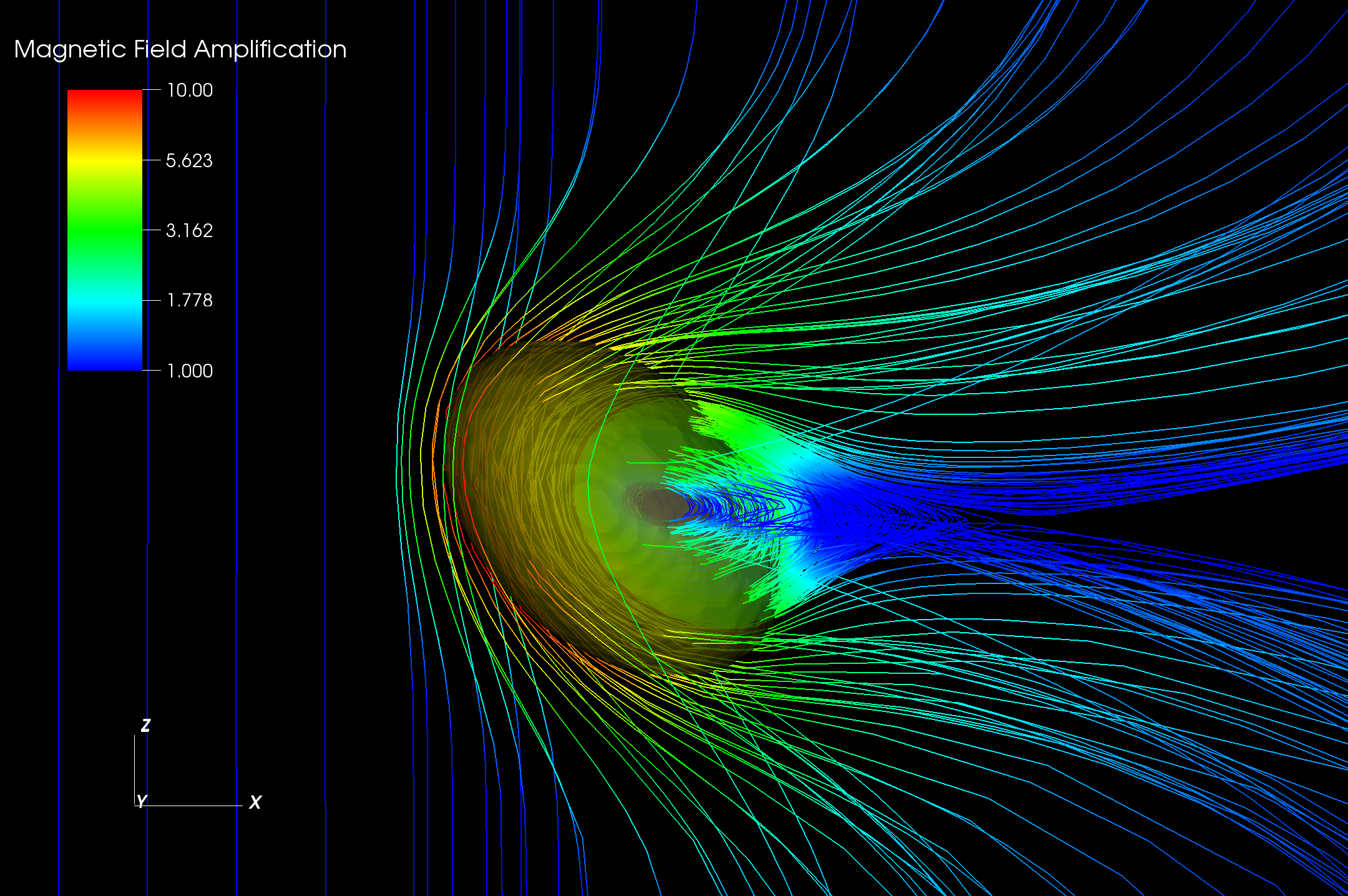}
\caption{Magnetic field lines colored by the (logscale) amplification relative to the solar wind field (100 nT) that arise from simulating the early solar wind overrunning the surface of one of our the (golden colored) bi-axial ellipsoid cases in Fig. \ref{fig:setupB}a, G3(55,45). The asteroid's surface is shown in translucent gray.}
\label{fig:lines}
\end{figure}

\begin{table*}
 \caption{Lower order magnetic moments of all cases (see Table \ref{tab:cases} for description). P cases have (16) Psych's $98 \degree$ orbital inclination. G cases show the 7 possible unit Cartesian coordinate orderings on an octant. $Q_{10}$ and $Q_{20}$ are the magnetic dipole and quadrupole moments in spherical coordinates (see Eqn. \ref{eqn:Vr});  $\mathrm{m_\rho}$, $\mathrm{m_\phi}$, and $\mathrm{m_z}$ are the dipole moments in cylindrical coordinates.  The spin aligned global asteroid magnetic dipole moment is $\mathrm{M_p=m_z}$.} 
 \centering
\begin{tabular}{crrrrrrr}
\multirow{3}[1]{*}{\textbf{Key}} & \multicolumn{2}{c}{\textbf{Spin Axis Angles}} & \multicolumn{2}{c}{\textbf{Spherical Moments}} & \multicolumn{3}{c}{\textbf{Cylindrical Dipole Moments}} \\
  \,    & $ \theta $ & $\varphi$ & $\mathrm{Q}_{10} $ & $ \mathrm{Q}_{20} $ & $\mathrm{m_\rho}$ & $\mathrm{m_\phi}$ & $\mathrm{m_z=M_p}$ \\
  \,    & ($\degree$) & ($\degree$) & (nT) & (nT) & \multicolumn{3}{c}{( $\times 10^{12} \, \mathrm{A \, m}^{2}$ )} \\
\toprule
P1 & 98 & 0 & -12.51 & -2.07 & 14.48 & -0.06 & -171.03 \\
P2 & 98 & 45 & -15.18 & -3.10 & 25.94 & -1262.90 & -207.63 \\
P3 & 98 & 90 & -19.49 & 0 & -0.01 & -1808.25 & -266.56 \\
P4 & 98 & 135 & -15.19 & 3.01 & -25.35 & -1263.07 & -207.69 \\
P5 & 98 & 180 & -12.50 & 2.07 & -14.66 & 0.05 & -170.94 \\
P6 & 98 & 225 & -15.18 & 3.01 & -25.17 & 1263.02 & -207.67 \\
P7 & 98 & 270 & -19.49 & 0 & 0.01 & 1808.24 & -266.59 \\
P8 & 98 & 315 & -15.17 & -3.01 & 25.50 & 1262.54 & -207.50 \\
PA & \multicolumn{2}{c}{Orbit Average} & -15.59 & -0.01 & 0.09 & -0.05 & -213.20 \\
\midrule
G1 & 90 & 90 & $-1 \times 10^{-3}$ & 0.01 & -0.17 & -1871.51 & -0.01 \\
G2 & 90 & 45 & $-4 \times 10^{-3}$ & -0.01 & 0.16 & -1289.66 & -0.06 \\
G3 & 54.74 & 45 & 59.30 & 9.98 & -76.43 & -873.87 & 810.99 \\
G4 & 45 & 90 & 77.65 & -0.01 & 0.12 & -954.64 & 1062.02 \\
G5 & 90 & 0 & 0.02 & -0.01 & 0.17 & 0.52 & 0.33 \\
G6 & 45 & 0 & 64.63 & 10.03 & -74.85 & 0.20 & 883.90 \\
G7 & 0 & 0 & 95.75 & -0.02 & 0.28 & 2.01 & 1309.49 \\
\midrule
CP* & \multicolumn{2}{c}{Dynamo (parent)} & 49.00 & -3.98 & 31.56 & $0$ & 670.16 \\
CB* & \multicolumn{2}{c}{Dynamo (body)} & $-5 \times 10^{-4}$ & $0$ & $0$ & $0$ & $-7 \times 10^{-3}$ \\
\bottomrule
\multicolumn{8}{l}{\footnotesize *The magnetic moments scale linearly to first order with the assumed core dynamo dipole moment } \\
\multicolumn{8}{l}{\footnotesize *We use parent body (CP) / internal dynamo (CB) values of: $ \mathrm{M_{CP}} = 2.25 \times 10^{18} \mathrm{A \, m}^{2}$ and $ \mathrm{M_{CB}} = 10^{16} \mathrm{A \, m}^{2}$ } \\
\multicolumn{8}{l}{\footnotesize Axial dipole ($Q_{10} = g_1^0$) and quadrupole ($Q_{20} = g_2^0$) are calculated using a radius of $a = 111 \text{ km}$ in Eqn. \ref{eqn:Vr}} \\
\end{tabular}%
\label{tab:all}
\end{table*}


\subsection{Solar Wind Induced Magnetization}

The synthetic global magnetic field, modeled by the superposition of local surface remanent magnetization dipoles, is represented by the colored lines in each panel of Fig. \ref{fig:xyAll}. Only the poloidal term contributes significantly to the global field, as the contributions from toroidal elements (like the $\mathrm{m_\phi}$ component in Table \ref{tab:all}) along each co-latitude cancel out upon integration.
In quantifying the predicted external field that the Psyche spacecraft will observe in its multiple orbits \citep{Elkins-Tanton+2020}, we take advantage of the axisymmetry that results from the spin-averaging procedure described in Section 2.4. The resulting magnetic field is expressible as the gradient of a magnetic scalar potential $V$, for which all terms vanish except $m=0$ modes, where $m$ is the azimuthal wave number \citep[e.g. IGRF Models]{Alken2021}.
We calculate the synthetic global field by a direct superposition of the potential from all $\sim$12,000 surface remanent dipoles. In this framework, these $m=0$ terms depend only on the surface remanent magnetization and the distance along the spin (z) axis from the center of each dipole, neither of which the averaging changes. The potential can be written in terms of the dipole ($Q_{10}$) and quadrupole ($Q_{20}$) contributions as

\begin{equation}\label{eqn:Vr}
   V(r, \theta) = \frac{a^3}{r^2} Q_{10} \cos \theta + \frac{a^4}{2 r^3} Q_{20} ( 3 \cos^2 \theta - 1 ).
\end{equation}
Since our model generates the predicted value of dipoles at the surface of (16) Psyche, we directly obtain the predicted net dipole moment from a volume weighted summation, and the quadrupole moment from numerical integration. These physical moments are then converted to the standard Gauss-Legendre coefficients ($Q_{10}=g_1^0, Q_{20}=g_2^0$) reported in Table \ref{tab:all} (see Section 2.4). More details can be found in Supplementary text S1 and \cite{Jackson1998ce}. The relative and absolute strengths of these moments can then be compared.

To systematically analyze the WIM results, we first isolate the effect of the azimuthal angle $\varphi$ (relative to the wind direction) by holding the polar angle $\theta$ constant. Comparing cases G5 ($\varphi=0\degree$), G2 ($\varphi=45\degree$), and G1 ($\varphi=90\degree$) in Table \ref{tab:all}, all with $\theta=90\degree$, we see the resulting axial dipole coefficient ($Q_{10}$) is negligible ($\approx 0$ nT) in all three. This demonstrates that for a spin axis perpendicular to the external magnetic field, the global dipole moment is minimal and insensitive to the $\varphi$ angle (relative to the wind).

In contrast, the polar angle $\theta$ (relative to the magnetic field) has a dominant effect on the magnetic topology. As $\theta$ decreases from $90\degree$ (e.g., G1, $Q_{10} \approx 0$ nT) to $45\degree$ (G4, $Q_{10} \approx 78$ nT) and finally to $0\degree$ (G7, $Q_{10} \approx 96$ nT), the global dipole moment increases to its maximum value. This confirms a strong $\cos \theta$-like dependence. The `P' cases for (16) Psyche's $98\degree$ tilt (close to, but substantively different than, $90\degree$) follow this trend, resulting in a consistent, small but non-zero dipole moment ($Q_{10} \approx -12$ to $-20$ nT) that depends weakly on the orbital phase ($\varphi$).

This analysis also highlights the importance of asymmetries in producing non-dipolar fields. Cases G3 ($\theta \approx 55\degree$) and G6 ($\theta=45\degree$), which are asymmetric with respect to both the wind and magnetic field axes, produce a significant axial quadrupole coefficient ($Q_{20} \approx 10$ nT). This is a very large component compared to the near-zero quadrupole in all other symmetric `G' cases and the orbit-averaged `PA' case. This confirms that these specific, asymmetric orientations (Fig. \ref{fig:xyAll}c, Fig. \ref{fig:lines}) are predicted to generate a non-dipolar field topology, a key prediction of the WIM model. Note that the magnetic field amplification in Fig. \ref{fig:lines} peaks in the compressed plasma pileup region within one grid zone outside of the asteroid's dayside surface, consistent with adaptive mesh refinement (AMR) resolution studies in \citep{Anand2021}.

If (16) Psyche cooled at its present-day inclination over a few orbital time periods, then an average of cases P1-8 is needed to predict the observed solar wind induced magnetic field. This is shown as case PA in Fig. \ref{fig:xyAll}(d) and Table \ref{tab:all}. Here the toroidal component of the surface dipole and any hemispherical asymmetry averages out, leaving a sizeable dipole term but a negligible quadrupole term.

\subsection{Core Dynamo Magnetization}
For the case of magnetization by a core dynamo (Fig. \ref{fig:xyAll}f) we first consider Runcorn’s theorem \citep{RUNCORN1975327}, which implies that the remanent dipole term of a spherical shell magnetized by an internal dipole source vanishes.
This happens because the radius of the shell is independent of the polar or azimuthal angle. (16) Psyche is an elliptical body, but we have found that for moderately elliptical axi-symmetric shells the dipole term is still
negligible. We quantify this in our CB (Core, Body) case: the resulting axial dipole coefficient $Q_{10}$ is effectively zero ($-5 \times 10^{-4}$ nT, see Table \ref{tab:all}), which is quantitatively consistent with Runcorn's theorem's prediction.
The result is a relatively weak multipolar field. While these results may be intuitive, modeling them was necessary to make a direct comparison to the wind induced magnetization cases.

In contrast, our `CP' (Core, Parent) case models (16) Psyche was a fragment of a larger body, for which we use the magnetic field of present-day Mercury as an analog (Fig. \ref{fig:xyAll}e). While asteroid-specific dynamo scaling laws are available \citep{Nimmo2009,Zhang2023}, the results are highly dependent on uncertain core/mantle parameters. We selected Mercury as an analog because it provides a clear, observationally-grounded lower bound for a dynamo and serves as a better compositional analog (metal-rich) for a potential parent body than other similarly-sized bodies (e.g., the icy moon Ganymede). To a first order approximation for an asteroid dynamo with dipole moment $\mathrm{M}$, our results in Table \ref{tab:all} would change as: $Q_{i0} = \mathrm{M} \, Q_{i0}/\mathrm{M_{CP \ or \ CB}} $. We have used: $ \mathrm{M_{CP}} = 2.25 \times 10^{18} \mathrm{A \, m}^{2}$, and $ \mathrm{M_{CB}} = 10^{16} \mathrm{A \, m}^{2}$.

Since it is statistically unlikely for the spin axis of the fragmented asteroid be aligned with the parent body’s core dynamo dipole field axis, the global field would generically look like a tilted dipole. The resulting global magnetic field is potentially the strongest of all cases explicitly modeled here. The surface remanent magnetization is not shown for this case as the entire volume
is the magnetized fragment that composes the asteroid.

\section{Discussion}
Here we discuss the key factors that influence our model predictions, the physical assumptions about the magnetization acquisition time, the challenges of inferring a remanent field in the presence of the present-day solar wind with (16) Psyche's tilt, and the scaling of our results with material properties.

\subsection{Acquisition Timescales and Model Assumptions}
The strength of any remanent field potentially exhibited by (16) Psyche will depend on the asteroid’s material properties, as well as the thermal and magnetic history of the surface layers over the acquisition time. A threshold value of $\rm{M_p}=2 \times 10^{14} \rm{A \, m}^{2}$ has been estimated for any field detection by the Psyche mission magnetometer team \citep{Weiss2023} by having a standoff distance outside the asteroid. This may need favorable solar illumination conditions \citep{orbit}, ideally an equinox, given (16) Psyche is triaxial and tilted at $98 \degree$. Here, we assume that the magnetic susceptibility is the same for both the core dynamo and the solar wind induced magnetization cases. 

The solar wind induced magnetization acquisition time in our case (i) is set by the post-collisional cooling rate. The fastest scenario would be instantaneous cooling following a slow ($\sim 2 \mathrm{\ km/s}$) collision with another asteroid \citep{Bland2014, Muxworthy2017}, but such magnetic acquisition is expected to be localized. 
However, following the pore-scale cooling framework of \cite{Bland2014}, it is plausible that impact-induced fracturing and brecciation by creation of a metal-silicate regolith layer \citep[e.g.][]{Consolmagno2008} could extend to the $\sim 10$ km depth needed to produce a magnetic signature visible from Orbit D ($M_\mathrm{p} > 2 \times 10^{14} \mathrm{\ A\ m}^{2}$) \citep{orbit}. Impacts can generate a porous mixture with $>40\%$ void space, breaking solid material into blocks of 10-100 m scale. In this initially porous structure, individual blocks cool on timescales of months to years, while the fracture network allows rapid heat loss through radiative transfer \citep{Ryan2022}. Iron-bearing minerals in asteroids also have higher thermal conductivity than pure silicates \citep[e.g.][]{Soini2020} which aids heat loss. This geometry may permit the entire fractured zone ($\sim 10$ km) to cool below magnetic blocking temperatures within the solar cycle time constraint, preserving the magnetic field orientation even as the material slowly compacts over subsequent millions of years. Cooling within a rotational period ($\sim 4$ hours) would be limited to depths less than 1 km, or result in highly localized or randomized magnetization if fragments cool individually and re-accrete following catastrophic disruption. For (16) Psyche specifically, the high orbital inclination (98$\degree$) results in nearly constant solar wind direction over extended periods, allowing magnetization without rapid directional changes. This permits cooling of fractured zones to intermediate depths (3-5 km) over timescales of months to years while maintaining coherent magnetization. Cases for which we assume rotational-period cooling are an intermediate calculation step toward the orbital-period cooling cases, and are worth exploring nonetheless. If the magnetic acquisition timescale exceeded several solar cycles, the solar wind induced fields would be highly multipolar, similar to the solar nebula cases (Supplementary Fig. S2) but much weaker, likely falling below the detection threshold from orbit.

Here, we largely follow this fast acquisition time cases. For an acquisition time less than an orbit time, the asteroid’s orbital position affects the observed field (see results from 8 different simulations corresponding to different locations in (16) Psyche’s orbit shown in Table \ref{tab:all}). Table \ref{tab:all} shows that the dipole moment ($\mathrm{m_z}$) depends most strongly on the axial tilt ($\theta$), with only a slight dependence on the orbital phase $(\varphi)$. This value of $\theta$ could have changed over time due to collisions, internal processes, tidal forces, secular perturbations, or thermal (such as Yarkovsky) effects. The last 3 effects are extremely slow. 
For the dynamo case, magnetization would be acquired as the parent body's mantle cooled through blocking temperatures over millions of years \citep[e.g.][]{tarduno2012}.
We assume the dynamo is relatively steady over the acquisition timescale.

We model the asteroid as a biaxial ellipsoid with spin axis along the shorter axis. This means an azimuthal symmetry such that spin averaging cancels $m>0$ modes exactly is not expected for a real asteroid. However, the $m=0$ modes of the 
magnetic scalar potential in equation \ref{eqn:Vr} do not depend on the averaging procedure. As such, we
expect triaxial and even non-uniform asteroids to have the same $m=0$ mode global field structure at Orbit C distances (190 km altitude) \citep{orbit}. Our model captures the dominant, low-order $m=0$ modes (dipole and quadrupole) of the remanent field; near-surface, high-order fields are not modeled as they would be undetectable at the planned orbital altitudes.

\subsection{Observational Signatures and Solar Wind Interaction}
To interpret magnetometer observations, it is useful to decompose the total measured magnetic field as $\mathbf{B} = \mathbf{B_1} + \mathbf{B_2} + \mathbf{B_R} + \mathbf{B_{sw}}$, where $\mathbf{B_1}$ represents perturbations from the plasma interaction (field line draping and pileup around the obstacle), and $\mathbf{B_2}$ is the induced field arising from electromagnetic induction (Faraday's law) in conductive layers responding to time-varying external fields, $\mathbf{B_R}$ is the internal remanent field (from WIM or dynamo magnetization), and $\mathbf{B_{sw}}$ is the background solar wind field. The first two terms ($\mathbf{B_1}$ and $\mathbf{B_2}$) are distinct physical processes that must be separated from the remanent field ($\mathbf{B_R}$) to constrain (16) Psyche's magnetization history.

The present day solar wind will interact with the remanent field and create a magnetopause outside (16) Psyche if $\rm{M_p} > 2 \times 10^{14} \rm{A \, m}^{-2}$. \cite{Fatemi2018} simulated the interaction of present day solar wind for both magnetized and unmagnetized (16) Psyche using a hybrid code that tracks ions separately from fluid electrons.
We can use their simulations to analytically assess the implications of the effect of the present day solar wind on our results. We use the standard Chapman-Ferraro standoff equation (Eq. \ref{stand}) \citep[e.g.][]{Greenstadt1971}, which provides a first-order approximation for the pressure balance between a dipole field and a supersonic wind. The standoff distance of the magnetopause for taking the highest magnitude (P3/P7) net dipole moment, $\rm{M_p} = 2.66 \times 10^{14} \rm{A \, m}^{2}$ from the P-cases in Table \ref{tab:all}, and present day solar wind values of velocity, $v=400\, \rm{ km\ s^{-1}}$, and density, $\rho=1 \ \rm{H}^{+} \rm{cm^{-3}}$, is 
\begin{equation}
    r_M \approx \sqrt[{6}]{\frac {\mu _{0}\rm{M_p}^{2}}{8\pi ^{2}\rho v^{2}}} \sim 127 \, \rm{km}
    \label{stand}
\end{equation} 
This calculated standoff distance is critically \textit{less than} (16) Psyche's largest semi-axis of 139 km. 
This implies that for even our strongest predicted `P' cases, the solar wind will directly impact the asteroid on its sunward-facing side, with a magnetopause forming only on the flanks. The anti-sunward wake region would instead contain a structured magnetotail. 
As \cite{Fatemi2018} show (see their Figures 1-2), a magnetized body (their Runs \#1 and \#2) still forms a structured magnetotail in the wake where the anti-sunward topology of the magnetic field will be preserved and can be observed by the spacecraft.

Looking at Figure 3 of \cite{Fatemi2018} and our previous simulations \citep{Anand2021}, it is clear that even an unmagnetized (16) Psyche can have a magnetic field exceeding the background solar wind field due to compressional amplification from the plasma interaction. 
This presents a significant observational challenge: a detected magnetic field is not, by itself, proof of a remanent field ($\mathbf{B_R}$) as it could also be a plasma interaction perturbation field ($\mathbf{B_1}$) from field line draping around even an unmagnetized obstacle.
A way to distinguish these scenarios lies in the magnetic topology. The unmagnetized, conductive case (Run \#3 in \cite{Fatemi2018}) produces a draped field and a simple plasma wake, whereas the magnetized cases (Runs \#1 and \#2) produce a ``magnetospheric-like structure'' with a distinct magnetotail.
We also note that magnetopause currents (e.g., Chapman-Ferraro currents) and other dynamic plasma effects not captured in our model would further complicate the observed field structure, reinforcing the need for a topological analysis in the shielded wake region.

In the anti-sunward wake, the spacecraft may be partially shielded from the plasma interaction field ($\mathbf{B_1} \rightarrow 0$), but the induced field from electromagnetic induction ($\mathbf{B_2}$) remains present globally. For a conducting body in a time-varying external field, the induced response depends on the magnetic diffusion timescale $\tau_{\rm diff} = \mu_0 \sigma R^2$ \citep[e.g.,][]{Jackson1998ce}. For the crustal resistivity adopted here ($1/\sigma \sim 100\;\Omega\,{\rm m}$, see Supplementary Section S2) and $R \sim 100$ km, $\tau_{\rm diff} \sim 10^2$ s. Since this is short compared to typical solar wind variability timescales (hours to days), (16) Psyche is in the resistive limit where induced currents decay rapidly. Furthermore, the perfect-conductor upper bound on the induced moment ($M_{\rm ind} \sim 2\pi R^3 B_{\rm sw}/\mu_0 \sim 10^{14} \rm{A \, m}^2$ for $B_{\rm sw} \sim 2$ nT) is comparable to or smaller than the remanent moments predicted here. The actual induced field will be further reduced below this limit. Nevertheless, separating these components through their distinct temporal signatures remains essential.

Separating the remanent field ($\mathbf{B_R}$) from induced ($\mathbf{B_2}$) and plasma interaction ($\mathbf{B_1}$) components also involves taking into account the spacecraft's motion through spatially varying fields. In the body-fixed reference frame, the remanent field should appear as a time-stationary component, while the induced field will correlate with solar wind variability. A frequency-domain analysis of the measured field, combined with regression against upstream solar wind conditions, can in principle isolate these components. The remanent field would be identifiable as the residual that is (i) constant in the body-fixed frame, (ii) uncorrelated with solar wind variations, and (iii) consistent with the topological predictions presented here. Multiple orbits under varying solar wind conditions will be essential for this decomposition.

If the magnetopause is compressed inside the body, as our calculation suggests, a careful multipole expansion of the structure of the magnetic field available on the anti-sunward (wake) side of the asteroid might be able to distinguish between the topology predictions of the case where only $\mathbf{B_1}$ and $\mathbf{B_2}$ contribute, the wind induced remanent cases (including the solar nebula), and the core dynamo case. The revolution period of (16) Psyche is $\sim 5$ years, so unless Orbit C coincides with either Equinox, the 100 days of both orbits C and D \citep{orbit} may be insufficient to measure our predicted seasonal variation of the magnetic field on the anti-sunward side from both hemispheres due to the 98$\degree$ axial tilt. This issue is compounded by the fact that orbit D is inclined at $ 160 \degree$ and will stay close to the equator.

\subsection{Material Properties and Field Strength Scaling}
Given that both the surface field of present day Mercury (which has the weakest known active dynamo), and typical surface fields in our wind models after post processing have annular regions of $B \approx 300 \, \rm{nT}$, it is reasonable to consider this as an example value being recorded.
We obtain an upper bound on the magnetic susceptibility based on iron-rich impact mixtures and other achondrites. However, potentially magnetized components of (16) Psyche might not be represented by mesosiderites, pallasites, iron meteorite, acapulcoites and lodranites. Instead, eucrite or diogenite meteorites \citep{McSween2013} might better represent relict crust or mantle components, respectively, and these have lower magnetic susceptibilities, $3 \lesssim \chi \lesssim 5$, and $ 0.004 \lesssim \chi_M \lesssim 0.4 $ \citep[e.g.][]{Macke2010}, with the remanent magnetic field scaling linearly as $ \chi_M $. For the lower $\chi_M$ cases, (16) Psyche will still record a field that might be measurable by the Psyche orbiter 
but it might be too weak to detect on the sunward side, and would need to be carefully quantified on the anti-sunward side. See supplementary information Figure S3 showing the effect of $d$ and $\chi_M$ on the dipole moment, as mentioned in Equation \ref{dipole}. 
Here we take a moderate estimate of the magnetic susceptibility $\chi_M \sim 1$.

In order for the closest approach of Psyche (75 km above the surface or $\sim 213$ km from the center) \citep{orbit} to be inside the magnetopause, such that a field can be detected and characterized under sunward illumination, 
a dipole moment $\rm{M_p} > 1.25 \times 10^{15}\rm{A \, m^2}$ is required from Equation (\ref{stand}).
For the magnetization depth $d =10$ km in Table \ref{tab:all}, this is only satisfied for the general WIM cases where $\theta \leq 17 \degree$ and the parent body core dynamo case would need a 2x stronger dynamo. For the WIM cases, the magnetic field observable by the orbiter will scale linearly with magnetization depth as per equation (\ref{dipole}). Fully mapping the field under sunward illumination at Orbit C will need $ \rm{M_p} > 2.6 \times 10^{15}\rm{A \, m^2}$.

As the internal structure of (16) Psyche is still unknown, we have also considered the possibility that it may have differentiated with a core that is orders of magnitude more conductive than the surrounding regions, but without an active dynamo. Using the framework of a previous study \citep{Anand2021}, we varied the size of the core to identify the core size that maximizes the solar wind induced magnetic field. We find that when the core is $75\%$ the radius of the body, the crust field increased by only about $ \sim 20 \% $ compared to a body having no core, so the influence on the solar WIM is not significant (see Supplementary Section S3 and Fig. S1).

\section{Conclusions and Future Directions}
Here we summarize our primary findings and the predicted field topological distinctions, discuss the concrete implications for the Psyche mission, the broader application of our models to other asteroids, and directions for future research.  

\subsection{Summary of Predicted Field Topologies}
We computed predictions for the axisymmetric, $m=0$, mode of dipole and quadrupole terms of the global magnetic field around (16) Psyche-like asteroids for solar WIM and core dynamo origins, and for different cooling timescales and asteroid spin orientations. We find that whether (16) Psyche acquired magnetization as a fragment of a larger parent body with an internal dynamo, versus external WIM can be determined by a combination of strength and topology of the field as long as sufficient data from both hemispheres is available.
In particular, our models predict the ratio of the dipole to the quadrupole coefficient is significantly non-zero for asymmetric WIM cases (e.g., G3), and the presence of these multipoles would further suggest external WIM. The parent body dynamo case also has the potential for a stronger remanent field
as potentially the entire volume of the asteroid is magnetized.
A strong global dipole field substantially
tilted to the asteroid’s spin axis would suggest magnetization as part of a larger parent body.
If instead of a coherent global field, the orbiter measures scattered pockets of local field, (i) for the parent dynamo case this could imply re-accretion of randomly oriented fragments to the main asteroid body or (ii) for the WIM case, this could imply magnetization from multiple smaller impacts. A complete absence of magnetic field could place an upper bound on the depth of the crust that can be magnetized by WIM, including the solar nebula wind.

\subsection{Observational Implications for the Psyche Mission}
A key challenge for the Psyche mission is distinguishing a weak remanent field (which our models predict to be $\sim$1--10 nT at orbital altitudes) from the background solar wind and the strong magnetic perturbations from the plasma interaction itself. As shown by hybrid simulations \citep{Fatemi2018}, even an unmagnetized but conductive (16) Psyche will generate perturbations of several nT through a draped field structure and a plasma wake (their Run \#3).
This implies that a simple detection of a magnetic field is not, by itself, proof of a remanent magnetization. The definitive test will be to distinguish the \textit{topology} and temporal variability of the field. 

Our remanent WIM and dynamo models predict stable, global dipole and quadrupole components that are constant in the body-fixed frame. As discussed in Section 4.2, the optimal strategy will be to analyze data from the anti-sunward (wake) region, where the asteroid shields the orbiter from the plasma interaction field ($\mathbf{B_1}$), allowing for a clearer measurement of the internal remanent field. However, even in the wake, the induced field ($\mathbf{B_2}$) from electromagnetic induction remains superimposed on any remanent field ($\mathbf{B_R}$). A multipole expansion of this wake-side vector data could potentially isolate the stable, non-dipolar ($Q_{20}$) or tilted dipole components we predict. Since both the induced field ($\mathbf{B_2}$) and remanent field ($\mathbf{B_R}$) can contribute dipolar and higher-order terms, characterizing the temporal variability of the observed topology is essential: multiple observations under different solar wind conditions will allow the time-varying induced component to be separated from the constant remanent field through regression against upstream solar wind measurements. This temporal separation must precede any multipole analysis, as the spherical harmonic decomposition is only meaningful for the isolated remanent component $\mathbf{B_R}$.
While a complete separation of internal and external fields is not possible without full hybrid modeling, a statistical analysis of the wake-side field over multiple orbits could potentially identify stable multipole components characteristic of remanent magnetization, as these would persist while solar wind fluctuations vary.

A complete absence of a detectable coherent magnetic field (a null detection) would also be a significant finding. It would place a stringent upper bound on the magnetization depth $d$ (see Eq. \ref{dipole}) and/or the magnetic susceptibility $\chi_M$ of the crust. This would imply that (16) Psyche either has a very thin or non-retentive regolith, or that it was never exposed to a sustained, coherent magnetizing field (WIM or dynamo) during its formation.

\subsection{Application to Other Asteroids}
The framework presented here can be applied to other M-class or metal-rich asteroids. Our three primary models (WIM, intact dynamo, fragment dynamo) represent fundamental formation pathways. For any given asteroid, if its spin axis, shape, and (critically) obliquity are known, our WIM model can be run to predict its specific spin-averaged remanent field. This provides a clear baseline to test against. For example, our finding that high-obliquity bodies (like Psyche) tend to acquire a stable, spin-aligned dipole (Case PA) while asymmetric orientations can be highly non-dipolar (Case G3) is a general prediction. 
The observational strategy mentioned in Section 4.2, i.e. utilizing temporal variability to separate the induced field ($\mathbf{B_2}$) from the constant remanent field ($\mathbf{B_R}$), is also applicable to other asteroids, though the magnetic diffusion timescale ($\tau_{\rm diff} = \mu_0 \sigma R^2$) will vary with each body's size and conductivity structure. 
This framework can thus be used to interpret magnetic field data from future missions to other differentiated or metal-rich bodies to help constrain their own unique formation histories.

\subsection{Future Directions}
There is currently no precise estimate of the epoch when (16) Psyche formed, as no meteorite samples are believed to have originated from the same parent body. The proto-solar turbulent nebular disk \citep[e.g.][]{Blackman2001} might not have cleared before (16) Psyche cooled and could have magnetized the asteroid via WIM from the nebular field. Asteroid cooling in the presence of the solar nebula could take many eddy crossing timescales in which case the interaction with many individual eddies with differently aligned magnetic fields would require averaging. Even conservatively accounting for field reversals during a cooling timescale lasting 100s of orbits, the conversion of ram pressure to magnetic energy would seemingly produce larger magnetic field strengths than either solar WIM or remanent magnetization from the core dynamo of a larger parent body. The field would then be highly multipolar and structurally distinct from solar WIM or dynamo sources (Supplementary Fig. S2). However, for large bodies, the bow shock reduces the effective ram pressure on the asteroid \citep{mai2018mag} and would weaken this nebular magnetic field amplification. More work is needed on this nebular WIM for a direct comparison with the two other field origin sources that we focused on. 

Finally, if the surface of (16) Psyche were directly magnetized by an impact which heated and cooled its outer layers rapidly compared to the rotation/tumbling rate of the asteroid, the outer layer would directly record the amplified magnetic field from the ambient wind. This would result in a large but highly localized field at the impact site and/or its antipode \citep{Crawford2020} and thereby induce very high order multipolar modes. High order modes could also arise if the crustal magnetic mineral composition is inhomogeneous, but this may be unlikely if (16) Psyche formed as a small fragment from the crust of a larger body. In that case, a highly asymmetric remanent field would suggest external WIM, whereas a symmetric field would suggest a dynamo in the progenitor body.

\section*{Open Research Section}
The code AstroBEAR used for our simulations can be cloned from GitHub \citep{astrobear}: \url{https://gitlab-public.circ.rochester.edu/astrobear/}

The user documentation can be found at: \url{https://bluehound2.circ.rochester.edu/astrobear/}

Specific problem modules for running the resistive MHD simulations, the MATLAB code for spin averaging and synthetic global magnetic field simulation, and derived data from this study 
are archived in Zenodo \citep{Anand2025Zenodo}: \url{https://doi.org/10.5281/zenodo.15053355}. 

The GitHub repository \citep{16-psyche} is available at: \url{https://github.com/hisenbug/16-Psyche}.

\acknowledgments
The authors thank the anonymous reviewers for comments and feedback which significantly strengthened this manuscript. AA acknowledges a Horton Graduate Fellowship from the UR Laboratory for Laser Energetics (LLE). The authors thank the Center for Integrated Research Computing (CIRC) at the University of Rochester for providing computational resources and technical support. This material is based upon work supported by the Department of Energy [National Nuclear Security Administration] University of Rochester ``National Inertial Confinement Fusion Program'' under Award Number(s) DE-NA0004144. Additional support are provided from grants NSF PHY-2020249, NASA grant 80NSSC19K0510, NSF EAR-2051550. 

\subsection*{Conflict of Interest}
The authors declare no conflicts of interest relevant to this study.

%
\bibliography{agusample} 
%


%
%
%
%
%

\end{document}


%
%


\title{Supporting Information for ``Early solar wind and dynamo magnetic field topology predictions for (16) Psyche and other asteroids''}
%
%

%
%



\authors{Atma Anand\affil{1}, Jonathan Carroll-Nellenback\affil{1,3}, Eric G. ~Blackman\affil{1,3}, John A. ~Tarduno\affil{2,1,3}}


\affiliation{1}{Department of Physics and Astronomy, University of Rochester, Rochester NY 14627}
\affiliation{2}{Department of Earth and Environmental Sciences, University of Rochester, Rochester NY 14627}
\affiliation{3}{Laboratory for Laser Energetics, University of Rochester, Rochester, NY 14623, USA}

%
%

%

\begin{article}

%
%

\noindent\textbf{Contents of this file}
\begin{enumerate}
\item Text S1 to S3
\item Figure S1 to S3
\end{enumerate}

\noindent\textbf{Introduction}
This supplementary document contains additional information about the Gauss-Legendre polynomials used for the multipole expansion, an explanation for the value of asteroid resistivity used, the effect of core size with a varying radius as the free parameter on the wind induced amplification, a figure showing the magnetic field lines from one of our AstroBEAR simulations, and a contour plot showing the asteroid's net magnetic dipole moment value for different magnetic susceptibilities and magnetization depth.



%

\subsection*{S1. Gauss-Legendre Polynomial Expansion}
For a region of space devoid of source terms of the magnetic field (free electric currents), the Ampere law reduces to $ \nabla \times \mathbf{B} = 0 $. The magnetic field is conservative, and can then be written as the gradient of a scalar potential, as the curl of a gradient is 0. We are generating synthetic magnetic field predictions for what the Psyche spacecraft will observe in its multiple orbits \citep{Elkins-Tanton+2020}, and one way to find the multipole expansion of the global magnetic field is non-linear curve fitting of the global field data at the distance ($a$ in Equation \ref{eqn:V}) from 16 Psyche's center corresponding to the height of the closest orbit of the flyer. The magnetometer data will likely be of sufficient resolution to  generate Gauss-Legendre polynomials for the multipole expansion of the asteroid's field up to at least the first few moments, similar to how the IGRF model \citep{Alken2021} is used to reconstruct Earth's field.

The magnetic scalar potential $V$ is given by
\begin{equation}\label{eqn:V}
    V(r, \theta, \phi) = a \sum_{n=1}^{N} \frac{a^{n+1}}{r} \sum_{m=0}^{n} \left[ g_n^m \cos(m \phi) + h_n^m \sin(m \phi) \right] P_n^m (\cos \theta)
\end{equation}

\noindent where, $P_n^m$ is the associated Legendre Polynomial,  $N$ is the order of our expansion (2 here), $ g_n^m $ and $ h_n^m $ are the Gauss-Legendre polynomial coefficients, $r, \theta, \phi$ are the spherical coordinates of a given point  centered on the asteroid. Given our fixed initial solar wind and magnetic field orientations, and an a biaxial ellipsoid that is axisymmetric about the rotational axis, the asymmetric $m>0$ modes vanish, and the $h_n^m$ coefficients are irrelevant,  leaving only $m=0$ modes. The single coefficient ($g_1^0 = Q_{10}$) then quantifies  the dipole term, and the single coefficient ($g_2^0 = Q_{20}$) quantifies the \textit{entire} Quadrupole tensor. 
After substituting the standard Legendre polynomials, this then reduces Eqn. \ref{eqn:V} to 
\begin{equation}\label{eqn:Vr}
	V(r, \theta, \phi) = \frac{a^3}{r^2} Q_{10} \cos \theta + \frac{a^4}{2 r^3} Q_{20} ( 3 \cos^2 \theta - 1 ).
\end{equation}

Since our model predicts values of dipole elements at the surface of 16 Psyche, we directly obtain the net dipole moment from a volume weighted summation of each element, and the quadruple moment from  numerically integrating them. The procedure is outlined in Section 4.1 of \cite{Jackson1998ce}, which also describes how to convert the moments from Cartesian to spherical coordinates. Table 1 of our main text lists the moments in  cylindrical coordinates which identifies some of the resulting surface dipole field patterns. The $z$ component differs from the spherical $z$ component by the constant factor introduced during conversion ($ \sqrt{\frac{4 \pi}{3}}$) and is the same in Cartesian coordinates. However, the cylindrical moments, $ m_\rho$, and $m_\varphi$, in the table are not corrected for analogous factors of order unity, but do not affect the qualitative interpretation. 

\subsection*{S2. Effective Resistivity of Surface Minerals}
The precise electrical resistivity of (16) Psyche's surface is unknown and depends on its composition and internal structure, for which there are multiple possibilities based on currently available data. As the effective resistivity of a mineral mixture (for instance, a pyroxene matrix with iron inclusions) is not readily available in the literature, we ran finite element mesh (FEM) simulations in AstroBEAR to determine the effective resistivity for several plausible compositions.These simulations confirmed that for all likely candidates, the electrical resistivity is low enough that two key conditions are met: (1) The physical magnetic Reynolds number ($R_M$) is high enough, such that $R_M \gg M\sqrt{\gamma\beta}$. This places the interaction in the high $R_M$ regime, where the limiting factor for compressive magnetic field amplification is the solar wind ram pressure, not the asteroid's intrinsic resistivity; (2) The magnetic diffusion timescale ($\tau \approx L^2 / \eta$) for the $\sim 10 \text{ km}$ crustal depths considered is much longer than the asteroid's rotational period ($\sim 4.2$ hours). 

Therefore, on the timescales relevant to our model, the magnetic field will pile up at the surface rather than rapidly diffusing through the crust. This means the exact value of the resistivity has a negligible effect on our simulation results, and we do not explore it further.

\subsection*{S3. Effect of Core Size on External Wind Field Amplification}
As an extension of our previous study, we varied the fraction of conducting core radius present inside a spherical asteroid while keeping all other parameters the same. The crust was given a resistivity of $ 100 \, \Omega m $, which is the order of magnitude resistivity we obtained from our simulations in the previous section, S2. The core was given a resistivity $ 10^6 $ times lower than the crust, but for the time scales our simulation runs it practically acts as a superconductor (solid iron has a resistivity of $ 10^{-6} \Omega m $). We considered 6 cases so as to obtain the radius at which the core size is so large that it begins to reduces the peak magnetic field amplification.

\begin{figure}
\noindent\includegraphics[width=\textwidth]{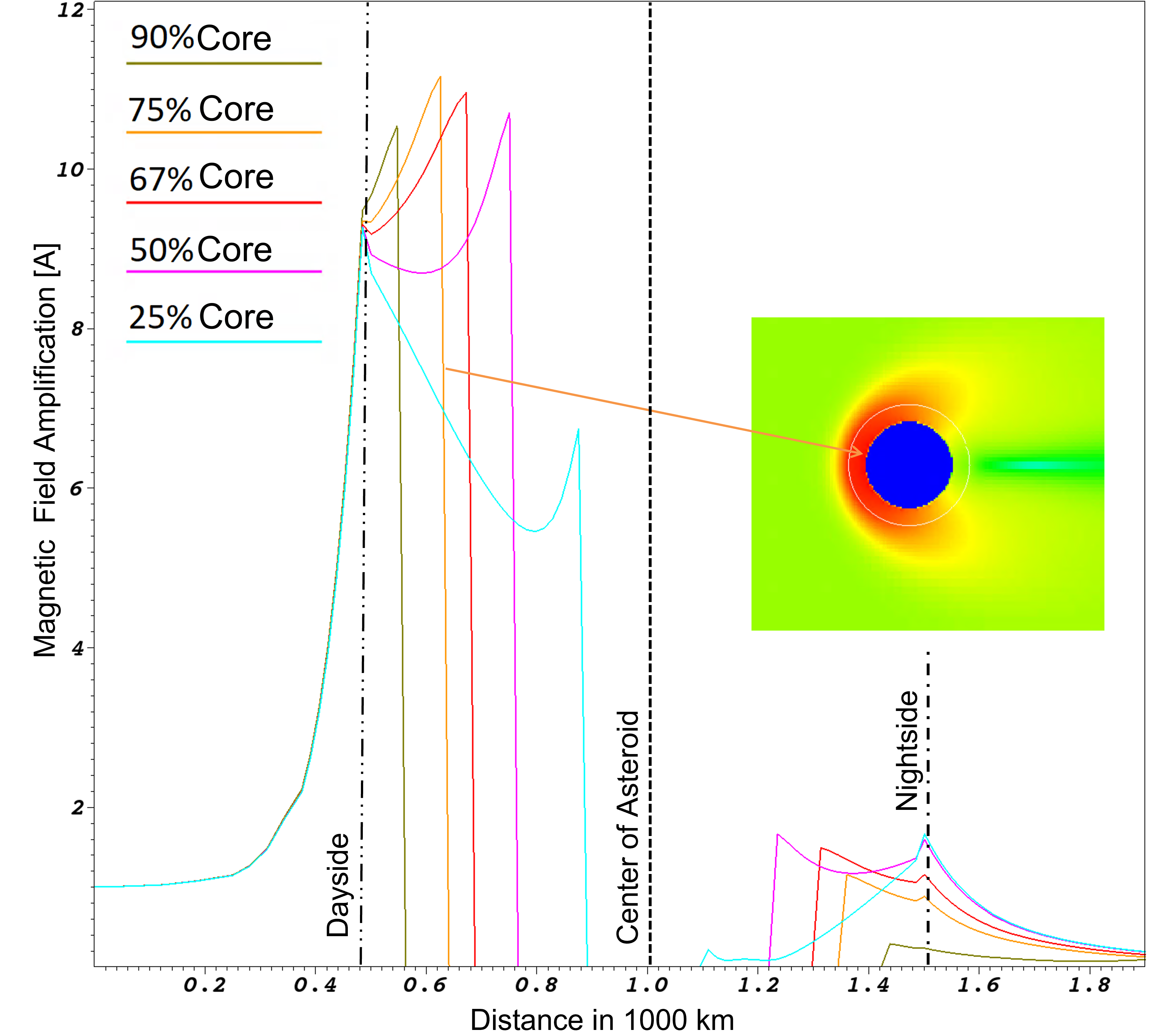}
\caption{Lineout of magnetic field amplification along the direction of solar wind flow passing the center of asteroid vs the distance starting 1 Radii ($=500$ km) in front of the asteroid. Solid lines show the amplification for models with a (practically) infinitely conducting core and an outer layer of resistivity of 100 $ \Omega m$. Inset shows magnetic field in a blue (low) to red (high) hot scale for the $ 75\% $ core case.}
\label{fig:cores}
\end{figure}

Fig. \ref{fig:cores} shows the lineout of magnetic field amplification for cores of various sizes for a spherical body.






%
%


%
%
%
%
%
\newcommand{\newblock}{}

\bibliography{agusample}

%
%
%
%
%

%
%
\end{article}
\clearpage


%

\begin{figure}
\noindent\includegraphics[width= \textwidth]{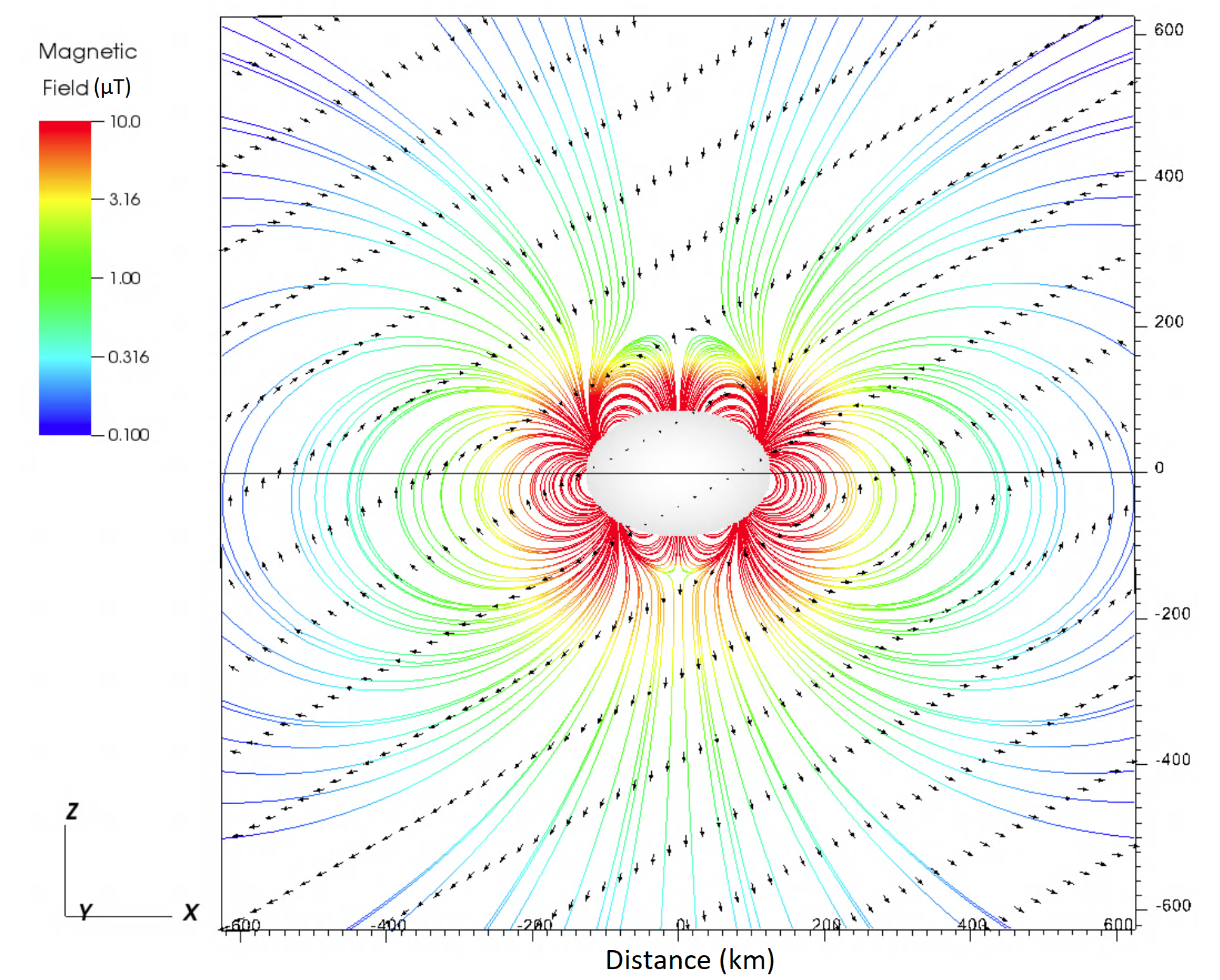}
\caption{\textbf{Solar nebula case:} Synthetic remanent magnetic field lines colored by magnitude and magnetic field direction (gray arrows) in the $y=0$ plane. The $z=0$ line highlights the North-South asymmetry. This case represents magnetization during exposure to the solar nebula over $10^5$--$10^6$ years, with the asteroid experiencing 1000 turbulent eddy crossings that randomize field orientation. The resulting topology is highly multipolar. Note that the displayed field strength assumes near full conversion of nebular ram pressure to magnetic pressure, which overestimates the actual field magnitude due to bow shock effects discussed in Section 5 and \cite{mai2018mag}.}
\label{fig:nebula}
\end{figure}

\begin{figure}
\noindent\includegraphics[width= \textwidth]{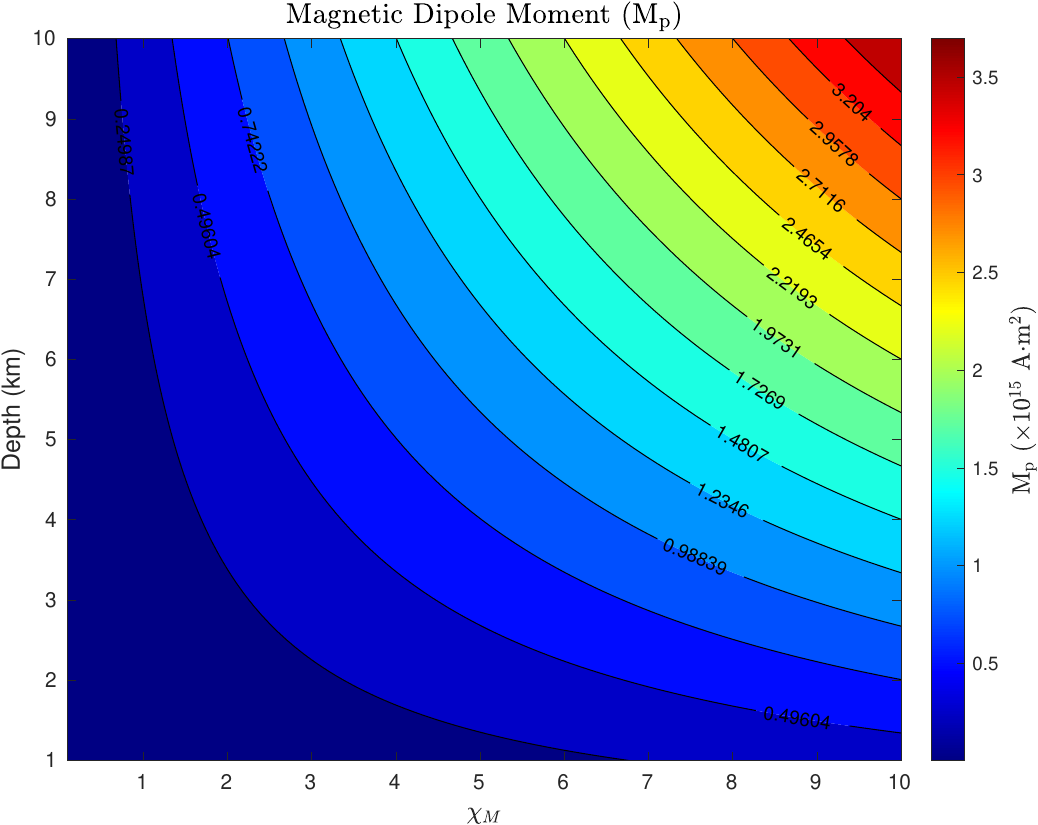}
\caption{Plot of the net Magnetic Field Dipole Moment of (16) Psyche assuming the seed field is 300 nT throughout the whole magnetization depth, and the radius is uniformly 111 km. The contours shows the degeneracy between the magnetic susceptibility ($\chi_m$) and magnetization depth ($d$). The point $\chi_m=4, \ d=10$ km in the main paper Fig. 2 and Table 1, lies along the contour $\rm{M_p \sim 1.25 \times 10^{15} \rm{A \ m}^2} $, which is also the minimum $\rm{M_p}$ needed to have the magnetopause above the orbiter's Orbit D.}
\label{fig:dipole}
\end{figure}